\pdfoutput=1
\documentclass[11pt]{article}

\usepackage[final]{acl}
\usepackage{multirow,booktabs}
\usepackage{times}
\usepackage{latexsym}
\usepackage{amsmath}
\usepackage{amssymb}
\usepackage{graphicx}
\usepackage{makecell}
\usepackage[most]{tcolorbox}
\usepackage{xcolor}

\usepackage[T1]{fontenc}
\usepackage[utf8]{inputenc}

\usepackage{microtype}

\usepackage{inconsolata}

\usepackage{graphicx}

\title{PKAG-DDI: Pairwise Knowledge-Augmented Language Model for Drug-Drug Interaction Event Text Generation}

\author{
 \textbf{Ziyan Wang\textsuperscript{1}\thanks{Equal contribution.}},
 \textbf{Zhankun Xiong\textsuperscript{1}$^{*}$},
 \textbf{Feng Huang\textsuperscript{1}},
 \textbf{Wen Zhang\textsuperscript{1}\textsuperscript{2}\textsuperscript{3}\thanks{Corresponding author.}},
\\
 \textsuperscript{1}College of Informatics, Huazhong Agricultural University, Wuhan, China\\
 \textsuperscript{2}Hubei Key Laboratory of Agricultural Bioinformatics,\\ Huazhong Agricultural University, Wuhan, China\\
 \textsuperscript{3}Engineering Research Center of Intelligent Technology for Agriculture,\\ Ministry of Education, Wuhan, China
\\
 \small{
 \{wangziyan,xiongzk,fhuang233\}@webmail.hzau.edu.cn, zhangwen@mail.hzau.edu.cn
 }
}

\begin{document}
\maketitle

\begin{abstract}
Drug-drug interactions (DDIs) arise when multiple drugs are administered concurrently. Accurately predicting the specific mechanisms underlying DDIs (named DDI events or DDIEs) is critical for the safe clinical use of drugs. DDIEs are typically represented as textual descriptions. However, most computational methods focus more on predicting the DDIE class label over generating human-readable natural language increasing clinicians' interpretation costs. Furthermore, current methods overlook the fact that each drug assumes distinct biological functions in a DDI, which, when used as input context, can enhance the understanding of the DDIE process and benefit DDIE generation by the language model (LM). In this work, we propose a novel pairwise knowledge-augmented generative method (termed PKAG-DDI) for DDIE text generation. It consists of a pairwise knowledge selector efficiently injecting structural information between drugs bidirectionally and simultaneously to select pairwise biological functions from the knowledge set, and a pairwise knowledge integration strategy that matches and integrates the selected biological functions into the LM. Experiments on two professional datasets show that PKAG-DDI outperforms existing methods in DDIE text generation, especially in challenging inductive scenarios, indicating its practicality and generalization\footnote{Our data and source code are available at https://github.com/wzy-Sarah/PKAG-DDI}.

\end{abstract}

\section{Introduction}
Unexpected drug-drug interactions (DDIs) may arise when people take multiple drugs simultaneously to treat complex diseases and potentially induce diverse pharmacokinetic and pharmacodynamic consequences, named DDI events (DDIEs). Predicting DDIEs holds significant importance for public health security and clinical research \cite{ryu2018deep}.

\begin{figure}[!t]\centering
	\includegraphics[width=\columnwidth]{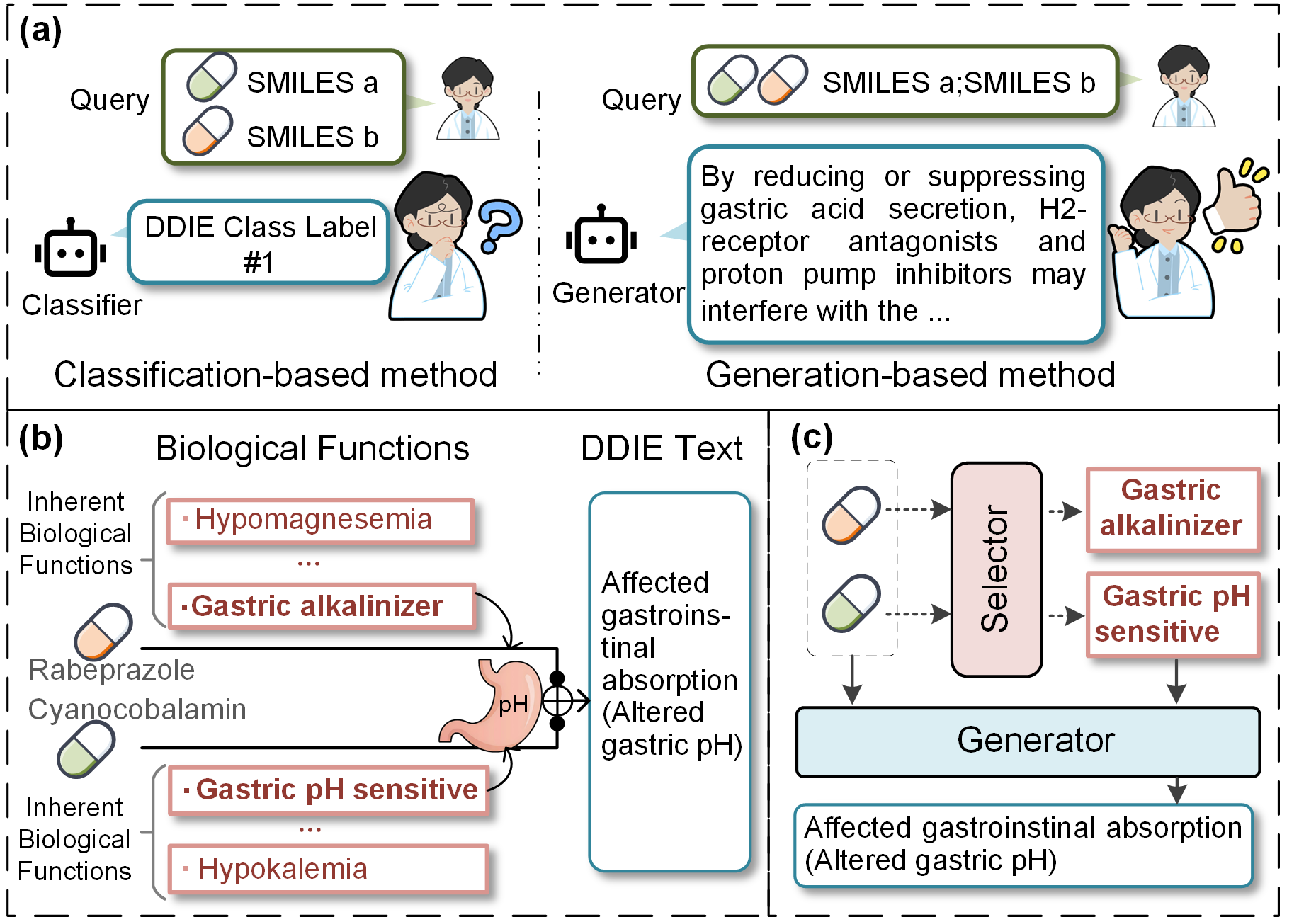}
	\caption{(a) The difference between classification-based and generation-based methods. (b) An example of a DDI. Every drug has several inherent biological functions. In a DDI, it has the most DDIE-relevant biological function. (c) When there are no prior biological functions, our method can use the selector to gather relevant biological functions to augment the DDIE prediction.}
    \label{FIG_1}
\end{figure}

Currently, most computational-based DDIE prediction methods \cite{ryu2018deep,xiong2023multi,knowddi} view the DDIE prediction as a classification task, which entails categorizing various DDIE texts into a finite number of predefined classes. On the one hand, predicting labels lacks intuitiveness, and predicting labels needs an extra label-to-text list to obtain corresponding DDIE information. On the other hand, the latest DDI databases, such as DDInter2.0 \cite{ddinter2.0}, provide more detailed DDIE texts, which are not readily classifiable. Therefore, as shown in Figure \ref{FIG_1} (a), in DDIE prediction, directly generating DDIE texts is a more natural way, highlighting the critical need to develop generation-based methods. 

%predicting labels needs an extra label-to-text list to obtain corresponding DDIE information, increasing the model application burden.

\begin{figure}[!t]\centering
	\includegraphics[width=0.98\columnwidth]{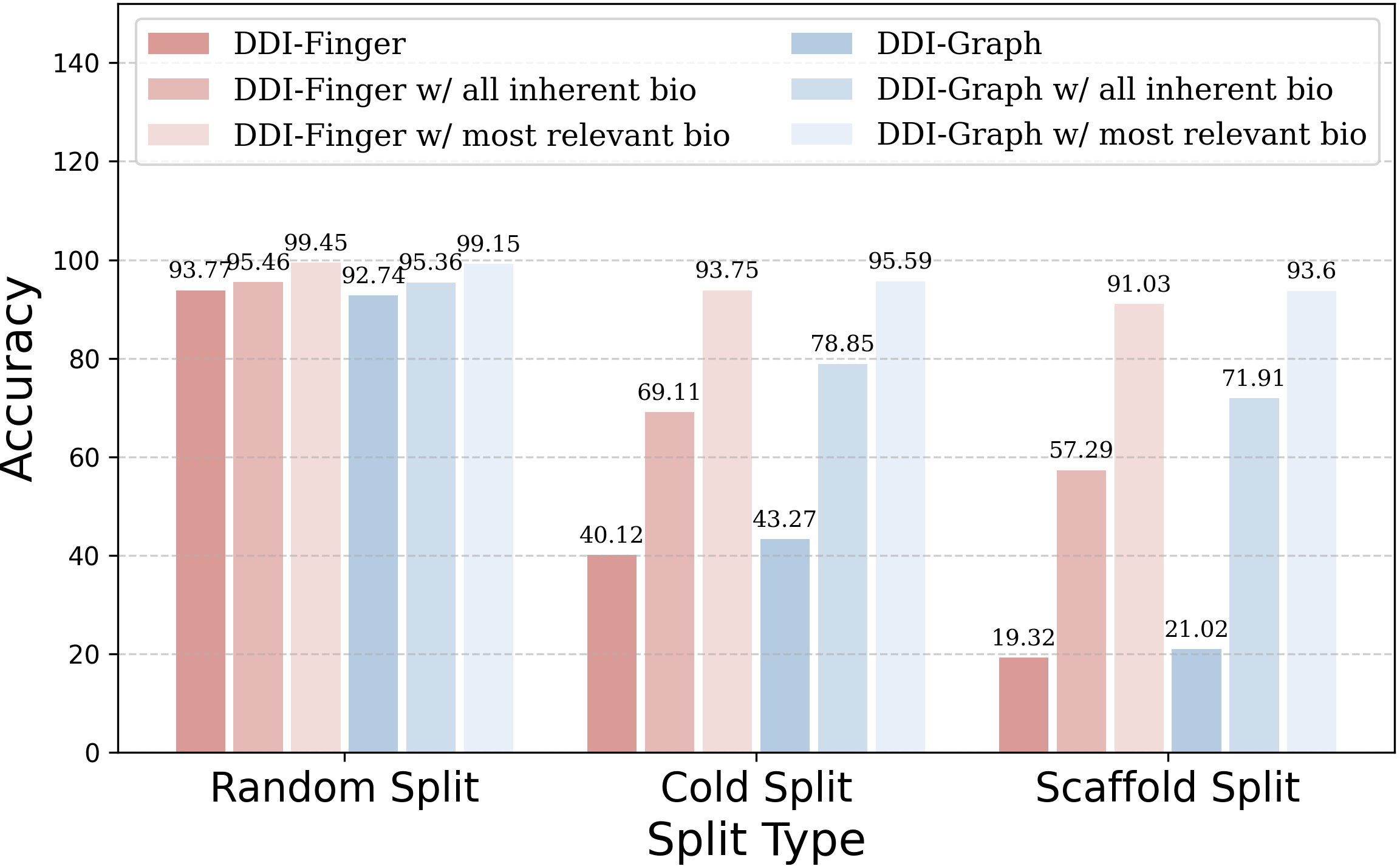}
	\caption{The multi-class classification performance of two basic models (DDI-Finger and DDI-Graph) on the MecDDI dataset. These models with the biological function as additional information (w/ all inherent bio, w/ most relevant bio) have remarkable improvement under three common DDIE prediction scenarios. More details are shown in Appendix \ref{sec:pre}.}
    \label{toy}
\end{figure}

Language models (LMs) have shown notable success in general text generation tasks \cite{brown2020language,luo_biogpt_2022}. A recent method MolTC \cite{2024-moltc} introduces an LM framework for molecular relational learning from molecular structure and tries to tackle the DDIE generation task. The strategy that infers the DDIE from the molecular structures of drugs may overlook the fact that each drug in a DDI has distinct biological functions (including drug mechanisms, activities, and categories) \cite{hu_mecddi_2023}. The biological functions can serve as a context to explicitly supplement and explain biological logical inference processes when DDI occurs. For example, as depicted in Figure \ref{FIG_1} (b), Rabeprazole, as \textit{gastric alkalinizer}\footnote{Herein, the italics refer to biological functions, which can explain, to some extent, why concurrent administration of Rabeprazole and Cyanocobalamin may impact gastrointestinal absorption.} can influence the activity of Cyanocobalamin, which is \textit{gastric pH sensitive}, thereby leading to gastrointestinal malabsorption caused by changes in gastric pH. The quantitative analysis shown in Figure \ref{toy} confirmed that the biological function information can significantly enhance the accuracy of DDIE prediction. Thus, incorporating biological functions holds promise for improving the capability of LM for DDIE text generation. However, biological functions are specialized knowledge that may not be readily accessible, potentially limiting widespread practical applications.

To tackle the above problem, we noticed retrieval-augmented generation (RAG), which has emerged as a popular knowledge augment technique for LM \cite{gao2023retrieval_survey,RAG_survey}. RAG leverages a retriever to extract several pieces of query-relevant knowledge from external databases when specialized knowledge is lacking and integrates the knowledge and original input to enhance the generation of LM. However, applying it to our task faces two challenges. (1) Every single drug has multiple inherent biological functions, and when interacting with another drug, it will have at least one most DDIE-relevant biological function (see Figure \ref{FIG_1} (b)). As Figure \ref{toy} shows that the most DDIE-relevant biological function is the key to the performance gains on DDIE prediction. According to this, the first challenge is how to model the mutual-conditioned DDIE-relevant biological function (named pairwise biological function) selection while maintaining efficiency. (2) Simply integrating all selected potential pairwise biological functions may introduce additional interference to LM generation, such as the noise from unmatched or irrelevant biological functions. How to design an effective biological function integration strategy is the second challenge.

In this work, we propose a novel pairwise knowledge-augmented generative method (PKAG-DDI) for DDIE generation, which selects the pairwise biological functions from the external knowledge set and takes them as context to enhance the DDIE generation of LM. Specifically, PKAG-DDI first designed a pairwise knowledge selector (PKS), which enables the mutual injection of molecular structural information between two drugs. Within the PKS, a reuse strategy is also introduced, sharing the weight of most components to reduce the computational burden. Then PKAG-DDI proposed a tailored pairwise knowledge integration strategy, which matches the selected biological functions and integrates them selectively into LM, boosting the LM's capability in generating DDIE texts.

Generally speaking, the main contributions of this paper are described as follows:

\begin{itemize}

\item We are the first to take the biological functions of drugs as the input context to augment LM for DDIE generation and propose a novel pairwise knowledge-augmented generative model (PKAG-DDI) for DDIE generation in real application scenarios, which can explicitly reveal the logical process underlying DDI occurrence.

\item We introduce a pairwise knowledge selector to efficiently select the pairwise DDIE-relevant biological functions from a knowledge set and a pairwise knowledge integration strategy to match and inject pairwise knowledge into LM for accurate DDIE generation.

\item The extensive experiments on two professional datasets show that PKAG-DDI outperforms existing methods in DDIE generation, especially in challenging inductive scenarios, 
indicating its practicality and generalization.
\end{itemize}

\section{Related Works}

\subsection{Drug-Drug Interaction Event Prediction}
Current DDIE prediction methods generally focus on classifying drug pair instances through label prediction, collectively called classification-based methods. Some of them construct a drug association network and subsequently employ GNNs to aggregate interaction information for DDIE prediction \cite{xiong2023multi,chen_muffin_2021,knowddi}, while others focus on developing efficient feature encoders (e.g., DNNs, GNNs) to learn drug pair representations from molecular identity information, including SMILES (Simplified Molecular Input Line Entry System) strings, fingerprints, 2D structures, and 3D structures for DDIE prediction \cite{nyamabo_ssiddi_2021,GMPNN,DSN-DDI,MSAN,3DGT-DDI}. Given that extracting identity information does not need global interaction information, the latter methods are more suitable for generalizing to the challenging inductive DDIE prediction scenario, where all test drugs are absent from the training set. Overall, the predictions of all classification-based methods lack intuitiveness and are restricted by predefined class boundaries.

Recently, language models (LMs) have gained widespread adoption and offer novel perspectives for diverse biomedical applications, such as MolT5 \cite{molt5} and MolTC \cite{2024-moltc}, which are devoted to translating biochemical language into readable natural language. In particular, MolTC leverages a cross-modal adapter combined with a pre-trained language model (LM) for molecular interaction prediction. However, these methods do not explicitly capture the individual drug's biological function within DDI.

\begin{figure*}[!t]\centering
	\includegraphics[width=0.98\textwidth]{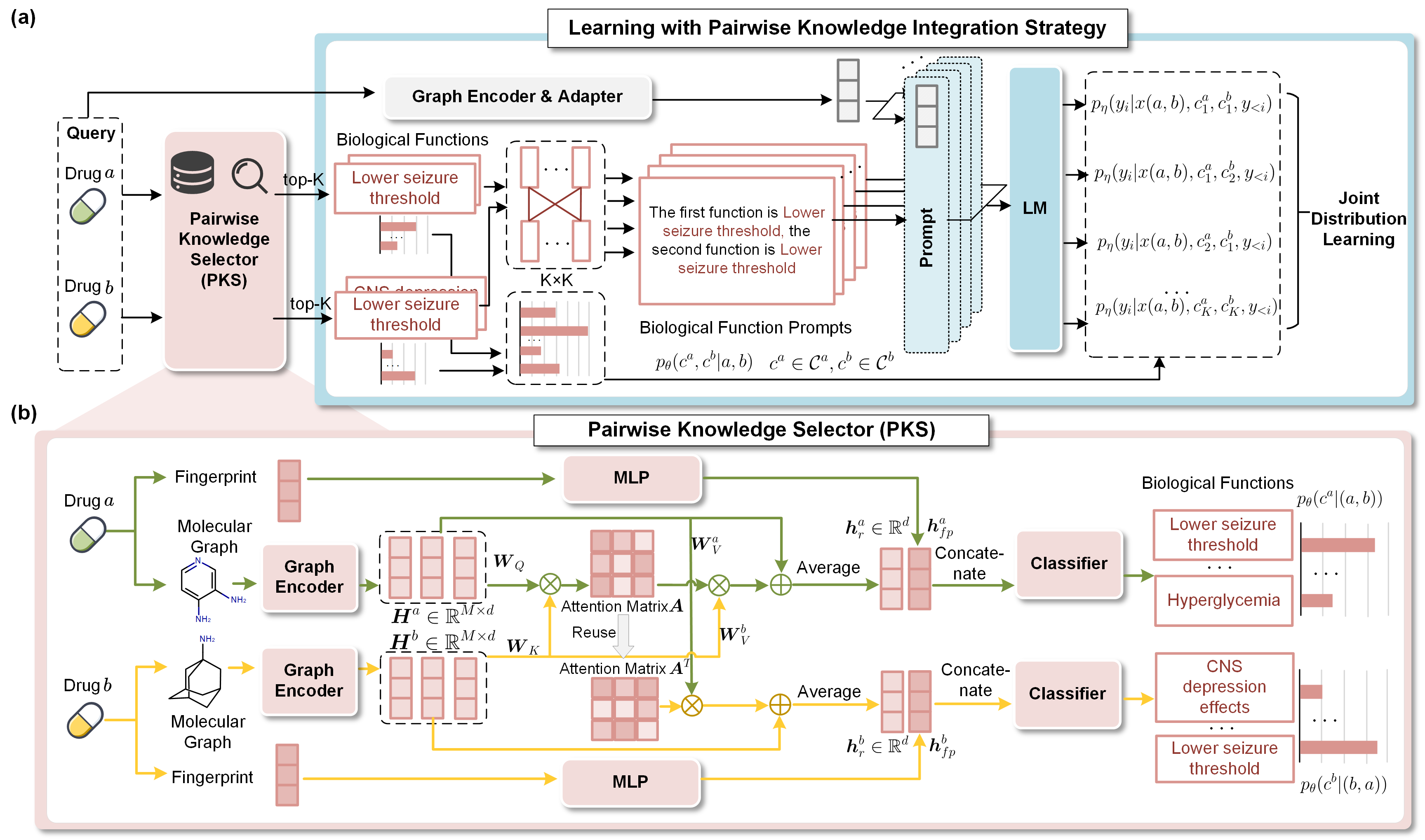}
	\caption{The overall framework of PKAG-DDI. (a) is the overall framework of PKAG-DDI. It firstly uses a PKS to select top-$K$ potential biological functions of each drug from the knowledge base and then match all possible pairwise biological functions and corresponding probability scores. Afterward, the matched pairwise biological functions together with the query information are put into LM for joint distribution learning by marginalizing all $K\times K$ pairwise biological functions. (b) PKS predicts $K$ biological functions of each drug. }
    \label{overview}
\end{figure*}

\subsection{Retrieval-Augmented Generation}

Retrieval-augmented generation (RAG) \cite{lewis2020retrieval} invokes a retriever to search and extract input query-relevant domain knowledge (such as facts or documents in a corpus) from external sets, and a generator that leverages the query alongside the knowledge to augment the generation \cite{gao2023retrieval_survey,RAG_survey}. Dense retrieval \cite{karpukhin-etal-2020-dense,lin2023fine,izacard_atlas_2024} embeds queries and external knowledge into continuous vector spaces and calculates relevant scores between them to get a ranked retrieved knowledge list, and can meet our needs for cross-modal retrieval (i.e., molecules to biological functions). However, since biological functions are domain-specific phrases and their scale is relatively small compared to large corpora, we streamline the retrieval process to a biological function selection/prediction task in our work, thereby reducing the computational burden, which voids the need for knowledge encoding knowledge by Bag of Words or BERT \cite{singh_end--end_2021}. Moreover, existing retrieval methods are not directly applicable to the pairwise retrieval objects in our task, necessitating further design.

% However, the core technologies in dense retrievals, such as encoding knowledge by Bag of Words or BERT \cite{singh_end--end_2021}, are overburdened for our biological function knowledge because the biological functions are specialized phrases and the scale of biological function is relatively small. Therefore, note that we streamline the retrieval process to a biological function selection/prediction task in our work to reduce the computational burden.

In addition, there are two commonly used input integration strategies for injecting the top-$K$ knowledge texts into generation. The first is concatenating the input query and all retrieved knowledge into a single prompt for the generation model \cite{ram_-context_2023}. This strategy may confuse LMs with irrelevant information \cite{xu_recomp_2024}. The second is concatenating each top-$K$ knowledge to the input, respectively, and ensembling output probabilities from all $K$ knowledge \cite{lin2023ra,shi2024replug}. However, they cannot tackle our pairwise knowledge task.

\section{Methodology}

\subsection{Problem Formulation}

The problem we aim to solve consists of stages: pairwise knowledge selector (PKS) and pairwise knowledge-augmented LM (PKA-LM). For selecting, given a drug pair $a$ and $b$ and a biological function set $\mathcal{C}$, the selector with parameters $\theta$ is to model two distribution $p_{\theta}(c^{a}| (a,b))$ and $p_{\theta}(c^{b}| (b,a))$ over the biological functions for drug $a$ and drug $b$, respectively. Here, we denote the primary drug $a$ interacting with another drug $b$ as $(a, b)$ and vice versa $(b, a)$. Note that $(a, b)$ and $(b, a)$ have their respective biological functions but lead to the same DDIE. $c^{a}, c^{b} \in \mathcal{C}$. For DDIE generation, given the input pairwise drugs and their potential pairwise biological functions, the generator $p_{\eta}(y|a,b,c^{a},c^{b})$ with parameters $\eta$ is to generate DDIE text $y$.

\subsection{Pairwise Knowledge Selector (PKS)}

\paragraph{Model Architecture.} PKS with the parameters $\theta$ aims to model two distributions $p_{\theta}(c^{a}| (a,b))$ and $p_{\theta}(c^{b}| (b,a))$, which first learn drug representations of $a$ and $b$, respectively, and then use two classifiers to predict the pairwise biological functions. The process is shown in Figure \ref{overview} (b), with detailed descriptions provided below.

Given a drug $a$, we first convert its SMILES to two commonly used initial molecular modalities by an RDKit tool: a fingerprint \cite{glen2006circular} and a molecular graph (i.e., atoms as nodes and bonds as edges). The molecular graph is then input into a graph encoder \cite{wang2024zeroddi} to get the molecular node prototype representations $\boldsymbol{H}^{a} \in \mathbb{R}^{ M \times d}$, where $M$ is the predefined number of node prototype and $d$ is the dimension. Similarly, we get the $\boldsymbol{H}^{b} \in \mathbb{R}^{ M \times d}$ for drug $b$. Then, $\boldsymbol{H}^{b}$ is used as conditional information and injected into $\boldsymbol{H}^{a}$ as the following cross-attention:

\begin{equation}
    \boldsymbol{A} = \frac{\boldsymbol{H}^{a}\boldsymbol{W}_{Q} \cdot {\boldsymbol{H}^{b}\boldsymbol{W}_{K}}^{T}}{\sqrt{d}},
\end{equation}
  
\begin{equation}
\boldsymbol{H}_{r}^{a}=\boldsymbol{H}^{a}+ \lambda \cdot Softmax\left(\boldsymbol{A}\right)( \boldsymbol{H}^{b}\boldsymbol{W}_{V}^{b}),
\end{equation}
where $\boldsymbol{W}_{Q}$, $\boldsymbol{W}_{K}$ and $\boldsymbol{W}_{V}^{b}$ are the learnable projection matrices. In this work, $\boldsymbol{A}$ denotes the attention scores between the node prototype representations of drug $a$ and drug $b$. $\boldsymbol{H}_{r}^{a} \in \mathbb{R}^{M \times d}$ is the information-injected node prototype representation of drug $a$, which is calculated by residual connection with attention-transferred $\boldsymbol{H}^{b}\boldsymbol{W}^{b}_{V}$ of drug $b$. $Softmax$ is the softmax operator, $\lambda$ is a hyper-parameter controlling the information flow from drug $b$. After averaging the rows of $\boldsymbol{H}^{a}_{r}$, we get the drug representation $\boldsymbol{h}_{r}^{a} \in \mathbb{R}^{d}$ for drug $a$. Distinctively, considering that constructing the $p_{\theta}(c^{a}| (a,b))$ and $p_{\theta}(c^{b}| (b,a))$ respectively may lead to redundant computations of the graph encoder, we propose a reuse strategy: reusing node prototype representations $\boldsymbol{H}^{a}$ and the attention scores $\boldsymbol{A}$ for computing $\boldsymbol{H}^{b}_{r}$ simultaneously with $\boldsymbol{H}^{a}_{r}$. To be more specific, the $\boldsymbol{A}$ can model the importance of drug $b$, meanwhile the transposed $\boldsymbol{A}^{T}$ can model the importance of drug $a$, thus we inject the information of drug $a$ for drug $b$ by:
\begin{equation}
\boldsymbol{H}^{b}_{r}=\boldsymbol{H}^{b}+ \lambda \cdot Softmax( \boldsymbol{A}^{T})( \boldsymbol{H}^{a}{\boldsymbol{W}_{V}^{a}}),
\end{equation}
where $\boldsymbol{W}_{V}^{a}$ is the learnable projection matrix. In a similar vein, we can get the drug representation  $\boldsymbol{h}^{b}_{r} \in \mathbb{R}^{d}$ for drug $b$. 

Subsequently, the $\boldsymbol{h}^{a}_{r}$ is concatenated with the fingerprint $\boldsymbol{h}^{a}_{fp}$, which has been transformed by a Multi-Layer Perceptron (MLP). Then, an MLP classifier is used to calculate the logits of $(a,b)$ to $c^{a}$:
\begin{equation}
\begin{split}
r(c^{a}|(a,b))= MLP_{c^{a}}(\boldsymbol{h}^{a}_{r} || \boldsymbol{h}^{a}_{fp}).
\end{split}
\end{equation}
Similarly, we get $r(c^{b}|(b,a))$, where $r$ refers to the relevant scores between a drug pair and a biological function. After that, a $Softmax$ is used to obtain the probability, i.e., $p_{\theta}(c^{a}|(a,b))$ and $p_{\theta}(c^{b}|(b,a))$  among all biological functions in the set.

\paragraph{Training.} We employ two negative log-likelihood losses for learning the distribution of $p_{\theta}(c^{a}| (a,b))$ and $p_{\theta}(c^{b}| (b,a))$, simultaneously. Note that since a drug pair may have more than one DDIE-relevant biological function, to simplify the learning object, we use the BM25 \cite{BM25} to select the most similar pairwise biological function to the corresponding DDIE text as the gold biological function labels for training. The reason is illustrated in Appendix \ref{sec:goldbiofunction}.

\subsection{Learning with Pairwise Knowledge Integration Strategy}

\paragraph{Model Architecture.} Given drugs $a$ and $b$ as query input and their potential biological functions as condition, the generator $p_{\eta}(y| a,b,c^{a},c^{b})$ with parameters $\eta$ is to generate DDIE text $y$. In particular, we propose a pairwise knowledge integration strategy to inject biological function information into the LM effectively. Hereafter, we detail the model architecture.

To construct the query input of LM, given drug pair $a$ and $b$, we first use the molecular graph encoder and the graph-to-sequence adapter from MolTC \cite{2024-moltc} to encode the molecular graphs of drug pair $a$ and $b$ to molecular token embeddings in text space, i.e.,
$\boldsymbol{T}^{a} \in \mathbb{R}^{Q\times d_{t}}$ and $\boldsymbol{T}^{a} \in \mathbb{R}^{Q\times d_{t}}$ for drug $a$ and $b$ , respectively. $Q$ is the number of tokens and $d_{t}$ is the dimension. Consistent with \cite{2024-moltc}, we also input the SMILES tokens of the drug pair $\boldsymbol{S}^{a}$ and $\boldsymbol{S}^{b}$ to LM. Thus, the query input $x$ is formulated as:
\begin{equation}
\begin{split}
x=x(a,b)=Prompt(\boldsymbol{S}^{a}, \boldsymbol{S}^{b}, \boldsymbol{T}^{a},\boldsymbol{T}^{b}),
\end{split}
\end{equation}
where $Prompt$ is the prompt text with the slots for $\boldsymbol{S}^{a}$, $\boldsymbol{S}^{b}$, $\boldsymbol{T}^{a}$ and $\boldsymbol{T}^{b}$. 

To effectively and selectively integrate biological function information, we aim to model the target $p(y|a,b)$ by a joint probability distribution that marginalizes all latent pairwise biological functions $c^{a}$ and $c^{b}$ based on \cite{REALM} rather than directly input all biological functions to LM:
\begin{equation}
\begin{split}
   p(y|a,b)=\sum_{c^{a},c^{b} \in \mathcal{C}}p_{\theta}(c^{a},c^{b}| a,b) p_{\eta}(y|a,b,c^{a},c^{b})
\end{split}
\label{eq_5}
\end{equation}
where $p_{\theta}(c^{a},c^{b}| a,b)$ is the pairwise biological function distribution consisting of $p_{\theta}(c^{a}|(a,b))$ and $p_{\theta}(c^{b}|(b,a))$. $\mathcal{C}$ is the biological function set. We append a pairwise biological function $c^{a}$ and $c^{b}$ to the query input $x$ to obtain the entire input of LM.

\begin{tcolorbox}[
    colback=gray!5!white,      
    colframe=gray!95!black,   
    title={The Entire Input of LM},
    fonttitle=\bfseries       
]
% \textbf{Input:} \verb|<SMILES>| (i.e., $\boldsymbol{S}$ ); \verb|<GraEmb>|(i.e., $\boldsymbol{T}$ ); \verb|<Function>| (i.e., $c$ ) 

The first drug is <$\boldsymbol{S}^{a}$> <$\boldsymbol{T}^{a}$>, the function is <$c^{a}$>. The second drug is <$\boldsymbol{S}^{b}$> <$\boldsymbol{T}^{b}$>, the function is <$c^{b}$>.

\end{tcolorbox}

Given that marginalizing all latent pairwise biological functions is resource-intensive, we approximate the Equation \ref{eq_5} by assuming over pairwise top-$K$ potential biological functions with the highest probability under $p_{\theta}(c^{a}|(a,b))$ and $p_{\theta}(c^{b}|(b,a))$, thus the Equation \ref{eq_5} is reformulated as:
\begin{equation}
 \begin{aligned}
    p(y|a,b) \approx \sum_{\substack{c^{a} \in \mathcal{C}^{a}\\c^{b} \in \mathcal{C}^{b}}}p_{\theta}(c^{a},c^{b}| a,b)p_{\eta}(y|a,b,c^{a},c^{b})  
\end{aligned}
\label{eq7}
\end{equation}
where $\mathcal{C}^{a} \subset \mathcal{C}$ and $\mathcal{C}^{b} \subset \mathcal{C}$ are the top-$K$ biological functions of drug $a$ and drug $b$, respectively. We match all possible biological functions from $\mathcal{C}^{a}$ and $\mathcal{C}^{b}$ and obtain $K \times K$ pairwise biological functions. One pairwise biological function probability is defined as the joint $p^{*}_{\theta}(c^{a}|(a,b))$ among top-$K$ and $p^{*}_{\theta}(c^{b}|(b,a))$ among top-$K$:
\begin{equation}
\begin{aligned}
p_{\theta}(c^{a},c^{b}| a,b)= p_{\theta}^{*}(c^{a}|(a,b)) p^{*}_{\theta}(c^{b}|(b,a))
\end{aligned}
\end{equation}
\begin{equation}
p^{*}_{\theta}(c^{a}|(a,b))=\frac{\exp (r(c^{a}|(a,b))}{\sum_{c^{a} \in \mathcal{C}^{a}}\exp(r(c^{a}|(a,b))}.
\end{equation}
Similarly, we can get $p^{*}_{\theta}(c^{b}|(b,a))$. According to this, the $\sum_{\substack{c^{a} \in \mathcal{C}^{a}\\c^{b} \in \mathcal{C}^{b}}} p_{\theta}(c^{a},c^{b}| a,b)=1$. Inspired by the token generation method \cite{lewis2020retrieval}, the current generated token is not only based on previous $i-1$ tokens but also influenced by $p_{\theta}(c^{a},c^{b}|a,b)$. Thus, finally $p(y|a,b)$ that generates DDIE with length $L$ is turn as:
\begin{equation}
 \begin{aligned}
    \prod_{i}^{L} \sum_{\substack{c^{a} \in \mathcal{C}^{a}\\c^{b} \in \mathcal{C}^{b}}}p_{\theta}(c^{a},c^{b}| a,b)p_{\eta}(y_{i}|x(a,b),c^{a},c^{b},y_{<i}) 
\end{aligned}
\label{eq10}
\end{equation}

\paragraph{Training.} 
Given that the aim of our method is designed for a professional DDIE generation, we freeze the parameters of the graph encoder and utilize our DDIE data to fine-tune the adapter and LM (leverages the medium-sized Galactica$_{1.3B}$ \cite{taylor_galactica_2022}) based on the pre-trained parameters in \cite{2024-moltc}. We optimize the model by minimizing each target's negative marginal log-likelihood in Equation \ref{eq10}.

\paragraph{DDIE Prediction.}
Taking the query input $x$ and top $K \times K$ pairwise biological functions from PKS, the generator utilizes the beam decoder to generate $K \times K$ DDIE texts. We take the one with highest generation scores $\text{exp}(p_{\theta}(c^{a},c^{b}| a,b)p_{\eta}(y|a,b, c^{a},c^{b}))$ as the prediction.

\section{Experiments}

\subsection{Experimental Setup}
%\subsubsection{Datasets} 
\paragraph{Datasets.}

To extensively evaluate the predictive ability of models, we constructed two DDIE datasets from two professional DDI databases, MecDDI \cite{hu_mecddi_2023} and DDInter2.0 \cite{ddinter2.0}, respectively. \textbf{\textit{MecDDI}} is a database that provides biological functions for all the collected DDIs\footnote{https://mecddi.idrblab.net/}. We collected all DDIs from the MecDDI database and filtered out drugs lacking SMILES and finally obtained the MecDDI dataset, which contains 1,685 drugs, 1,061 types of biological functions, and 152,922 DDIs belonging to 103 types of DDIE. \textbf{\textit{DDInter2.0}} is a comprehensive DDI database \footnote{https://ddinter2.scbdd.com/}. Different from MecDDI having highly-summarized and countable DDIE, the DDIE descriptions in DDInter2.0 are more detailed and hard to summarize into countable classes directly, as shown in Appendix \ref{sec:Dataset}.  After our collection, filtration, and ensuring every DDI has the MecDDI-provided biological functions of drugs, the DDInter dataset, in our work, contains 1,683 drugs and 152,887 DDIs.

%\subsubsection{Baselines}
\paragraph{Baselines.}

We compare our method with the following two types of baselines:

\begin{itemize}
\item  \textbf{\textit{Classification-Based Methods}}: DeepDDI \cite{ryu2018deep} utilizes the structural similarity of drug pairs to predict the DDIE labels. GMPNN-CS \cite{GMPNN}, SSI-DDI \cite{nyamabo_ssiddi_2021}, SA-DDI \cite{saddi}, MSAN \cite{MSAN}, and DSN-DDI \cite{DSN-DDI} design different GNN-based encoders to learn representations of 2D molecular structures of drug pairs for DDIE prediction. 3DGT-DDI \cite{3DGT-DDI} encodes 3D structure information of drug molecules through a molecular conformation encoder and follows a CNN for DDIE prediction.

% The aforementioned methods were applied in cold-start experiments in their work.
\item \textbf{\textit{Generation-Based Methods}}: MolTC \cite{2024-moltc} takes the SMILES and the 2D structure of drug pairs as input and is supervised by molecular descriptions for pre-training. Herein, to compare with our method, we fine-tune all its parameters in its fine-tuning stage with our dataset. MolT5 \cite{molt5} inputs the SMILES of a drug molecule originally, which is modified by adding another drug molecule for tackling our DDIE generation task.
\end{itemize}

\begin{table*}[t]
\resizebox{\textwidth}{!}{
\begin{tabular}{lcccccccccccccc}
\toprule
 \multirow{2}{*}{\makecell{Model}}& \multirow{2}{*}{Dataset}  & \multicolumn{4}{c}{Random Split} & \multicolumn{4}{c}{Cold Start Split} & \multicolumn{4}{c}{Scaffold Split} \\ \cmidrule(l){3-6} \cmidrule(l){7-10} \cmidrule(l){11-14}
  & & B-2 & B-4 & METEOR  & R-L & 
  B-2 & B-4 & METEOR  & R-L & 
   B-2& B-4 & METEOR  & R-L       \\\midrule
 
 MolT5 & \multirow{4}{*}{MecDDI}& 93.51 & 92.71 & 94.09 & 93.59 & 62.82  &  58.05 & 64.29 & \underline{61.95} &47.78 & 41.97 & \underline{50.29}& \underline{46.45}\\
  MolTC  & & 96.02 &95.59 & 96.12 
  &96.02& 59.84 & 55.16 & 61.51  & 58.85 &44.88&39.33 & 46.65 &42.39\\
   PKAG-DDI  & & \underline{96.75} & \underline{96.39} &\underline{96.85} & \underline{96.72}& \underline{62.96} & \underline{58.71} & \underline{64.34} & 61.87 & \underline{48.30} & \underline{42.89}& 49.53   & 45.78 \\
  PKAG-DDI$^{*}$  & & \textbf{99.63}& \textbf{99.58}& \textbf{99.68} &\textbf{99.89} & \textbf{99.19} & \textbf{99.35}  & \textbf{99.21}& \textbf{99.58} & \textbf{99.17} & \textbf{99.29} & \textbf{99.19} & \textbf{99.46}\\ \midrule
  
  MolT5 & \multirow{4}{*}{\makecell{DDInter\\2.0}}&81.30 & 80.30 & 86.02& 86.38& 29.01 & 24.21 & 34.38  & 31.99 & 10.94 & 5.84  &18.40  & 15.96\\ 
  MolTC  & & 83.18 &  81.71&86.50 & 85.84 &37.34 &32.36 &40.71&39.47& 20.40& 14.20 &28.31&28.39\\
  PKAG-DDI & & \underline{92.39}& \textbf{91.54}& \underline{92.78}  &\underline{92.10} &\underline{44.18}& \underline{39.42}&\underline{46.13}& \underline{43.66}& \underline{22.85 }& \underline{16.23} &\underline{30.45}& \underline{28.66} \\ 
  PKAG-DDI$^{*}$ & & \textbf{92.43} & \underline{91.49} &\textbf{92.88}  & \textbf{92.17} & \textbf{56.35} & \textbf{52.10} &\textbf{57.49}   & \textbf{56.59} & \textbf{39.15} & \textbf{33.00}&\textbf{48.74} & \textbf{48.62}\\ 
  
\bottomrule
 
\end{tabular}}

\caption{ Results (in \%) of our method with generation-based baselines for DDIE generation on the MecDDI dataset and DDInter2.0 dataset under three different data split settings. The abbreviations are BLEU-2, BLEU-4, and ROUGE-L, respectively. The best and suboptimal results are highlighted in \textbf{bold} and in \underline{underline}, respectively. }
\label{baselines1}
\end{table*}

\begin{table}[]
\resizebox{\columnwidth}{!}{
\begin{tabular}{lcccccc}
\toprule
\multirow{2}{*}{Model}  & 

\multicolumn{2}{c}{Minor} & \multicolumn{2}{c}{Moderate} & \multicolumn{2}{c}{Major} \\ \cmidrule(l){2-3} \cmidrule(l){4-5}  \cmidrule(l){6-7} 
& \multicolumn{1}{c}{B-2}   & \multicolumn{1}{c}{R-L}     & \multicolumn{1}{c}{B-2}     &
\multicolumn{1}{c}{R-L} &
 \multicolumn{1}{c}{B-2} & \multicolumn{1}{c}{R-L}
 \\ \midrule
MolTC     &  75.15  &   79.09  & 83.89 &86.98 &82.39 & 83.89 \\
PKAG-DDI   & \textbf{85.86} &  \textbf{87.08}   & \textbf{93.66}&  \textbf{93.12} & \textbf{88.80} & \textbf{89.61}  \\  \bottomrule  
\end{tabular}}
\caption{The quality of generated text in different clinical risk levels.}
\label{risk}
\end{table}

\paragraph{Evaluation Protocols.}

In this work, we comprehensively measure methods under three DDI scenarios: \textbf{\textit{Random Split}} simulating transductive scenario means we randomly split the samples (drug pairs and corresponding DDIEs) in the dataset to training, validation, and testing sets by 7:1:2. Next are inductive scenarios: \textbf{\textit{Cold Start Split}} means we split drugs into seen drugs and unseen drugs in the ratio of 2:1, the drug pairs in the training set only involve seen drugs, the drug pair in the validation set is composed of seen drug and unseen drug, the drug pairs in the testing set only involve unseen drugs. \textbf{\textit{Scaffold Split}} is the same as Cold Start Split except for the difference that the seen and unseen drugs are split by molecular scaffold. All results are the means of 3 independent runs. We conduct separate training and evaluation processes for both datasets.

% DDInter$_{\text{unseen}}$ is used to test the generalization ability of methods.

% The discussion about clinical relevance of these is shown in Appendix \ref{relevance}.

\paragraph{Metrics}
For generation task, we employ BLEU-2, BLEU-4 \cite{papineni2002bleu}, ROUGE-L \cite{lin2004rouge}, and METEOR \cite{banerjee2005meteor} for quantitative analysis. For classification, we employ Accuracy and Macro-F1 to measure the results. Note that to qualitatively measure the performance of generation-based methods, we convert them to text classification, i.e., we first vectorize \cite{harris1954distributional} the predicted text and the texts of all DDIE classes, then calculate the cosine similarity between them. The most similar label is the prediction. The implementation details and hyper-parameters are shown in Appendix \ref{sec:config}

\subsection{Comparison with Baselines}

In this section, we evaluate the performance of our method using both textual generation and multi-class classification metrics. To measure the effectiveness of biological function in LM, we propose an upper-bound model, PKAG-DDI$^{*}$, which directly adds the pairwise gold biological function to the input prompt of the generator for the prediction of DDIE.

\paragraph{Evaluate the Generative Capacity.} We compare our methods with the generation-based methods on the MecDDI and DDInter2.0 datasets under three data splitting scenarios. The results are shown in Table \ref{baselines1}. Except for our PKAG-DDI$^{*}$, PKAG-DDI achieves optimal performance in almost all evaluation metrics in both datasets, indicating the remarkable and consistent superiority of PKAG-DDI in the generation of DDIE text. Moreover, we have the following observations: (1) Performance of all methods decreases when replacing the experiment data from the MecDDI to the DDInter2.0, highlighting the difficulty of generating more detailed and complex DDIE text on DDInter2.0. In particular, PKAG-DDI shows prominent advantages in DDInter2.0, which indicates that PKAG-DDI exhibits robust capabilities for long and complex DDIE text generation. (2) The superior performance of the upper-bound model PKAG-DDI* underscores the significant contribution of capturing biological functions to DDI text generation. Moreover, PKAG-DDI achieves comparable results with PKAG-DDI* in some scenarios, validating the effectiveness of our pairwise knowledge integration strategy.

To further evaluate the real-world applicability of the generated text, we categorize all DDIs into three clinical risk levels (i.e., Minor, Moderate, and Major). We compare our method with the suboptimal baseline (MolTC) on the Random Split set of DDInter 2.0, and the results are shown in Table \ref{risk}. The results indicate that in different clinical risk level scenarios, the generated text of our method has a significant advantage. In addition, performance at the Major level is higher than that at the Minor level, indicating our method's high value in clinical major risk assessment.

\begin{table}[]
\resizebox{0.5\textwidth}{!}{
\begin{tabular}{lccccccccccccc}

\toprule
\multirow{2}{*}{Model} & \multirow{2}{*}{\makecell{Model\\ Type}}  & \multicolumn{2}{c}{Random Split} & \multicolumn{2}{c}{Cold Start Split} & \multicolumn{2}{c}{Scaffold Split} \\ \cmidrule(l){3-4} \cmidrule(l){5-6} \cmidrule(l){7-8}
 &   &  ACC.  &  F1   &  ACC.  &  F1   &  ACC.  &  F1       \\ \midrule

DeepDDI & \multirow{7}{*}{\makecell{Classifi-\\cation\\ based}} &88.68 & 76.56  & 37.91 & 21.03   & \underline{20.10} &4.62    \\

GMPNN-CS &  & 68.89 & 40.62  & 26.96 & 12.50 & 18.89 & \textbf{5.58}    \\

SSI-DDI  &  &90.04 &80.19 & 36.00 &  22.04  &  16.54 &  4.24  \\

DSN-DDI&  &91.03  &81.09 & 39.83 &  23.63 &  19.55     &   4.77    \\

MSAN  &  & 93.53  & 76.52  & \underline{42.45} & 22.22  &14.30 &1.09
     \\
SA-DDI  &  & \textbf{95.29} & \textbf{90.04}   & 38.32 & 25.46 & 16.48 &4.11 
     \\ 
3DGT-DDI & &92.86
 & 82.23& 36.96 & 19.28  &   19.60    &    4.38      \\\midrule
   
MolT5   & \multirow{3}{*}{\makecell{Genera-\\tion\\ based}} &    90.78    &    81.81     &   40.94  & 24.46    &    19.46    & 4.80  \\

MolTC   &      &   94.12  &    87.51   &   40.41    &  \underline{27.46 }    &   18.11    &    4.80   \\ 

 PKAG-DDI & & \underline{95.05} & \underline{87.96}  & 
 \textbf{44.39}& \textbf{27.80}   & \textbf{21.97} &  \underline{5.54}    \\ 

\bottomrule  
\end{tabular}
} 
\caption{ Multi-class classification performance (in \%). ACC. is the abbreviation for Accuracy.}
\label{baselines2}
\end{table}

\paragraph{Evaluate the Classification Capacity.}
We further assess the quality of our method's generated DDIE text using classification metrics and compare its performance with generation-based methods and classification-based DDIE prediction baselines on the MecDDI dataset. The results are shown in Table \ref{baselines2}. Our proposed method, PKAG-DDI, still achieved competitive performances. On the one hand, this confirms that our proposed method does not sacrifice DDIE prediction accuracy for the sake of DDIE text generation. On the other hand, it demonstrates that DDIE text generation, as an emerging approach to DDIE prediction, holds significant potential and value for practical applications.

\begin{table}[]
\resizebox{\columnwidth}{!}{
\begin{tabular}{lllllll}
\toprule
\multirow{2}{*}{Model}  & 

\multicolumn{2}{c}{Random Split} & \multicolumn{2}{c}{Cold Start Split} & \multicolumn{2}{c}{Scaffold Split} \\ \cmidrule(l){2-3} \cmidrule(l){4-5}  \cmidrule(l){6-7} 
& \multicolumn{1}{c}{A.@2$\uparrow$}   & \multicolumn{1}{c}{Time$\downarrow$}     & \multicolumn{1}{c}{A.@2$\uparrow$}     &
\multicolumn{1}{c}{ Time$\downarrow$} &
 \multicolumn{1}{c}{ A.@2$\uparrow$} & \multicolumn{1}{c}{ Time$\downarrow$}
 \\ \midrule
PKR w/ BERT      &  90.38  &   170.8  & 46.28 &90.3 &26.12 &  95.7  \\
PKR w/ BoW   &  97.50  & 476.2 &  47.06&261.6 & \textbf{26.63}& 255.4 \\
PKS w/o Reuse   &       98.15   &   134.1       &51.52 & 71.7&23.88 & 62.7 \\
PKS   & \textbf{98.51} &  \textbf{65.0}   & \textbf{53.98}&  \textbf{41.9} & 25.80 & \textbf{35.2}  \\  \bottomrule  
\end{tabular}}
\caption{The classification performance of the PKS and the constructed comparison methods on the MecDDI dataset. A.@2 refers to top 2 Accuracy, and the Time refers to the wall clock time of inference.}
\label{retrieval}
\end{table}

\subsection{Efficiency Analysis of Pairwise Knowledge Selector (PKS)}

In this paper, we simplify the pairwise knowledge retriever to the selector, as mentioned in Related Work, and propose a reuse strategy. Thus, we assess the efficiency of PKS by comparing it with its variants, including dense retrieval variants with Bag of Word encoder and BERT encoder (dubbed PKR w/ BoW and PKR w/ BERT) and the variant without reuse strategy (PKS w/o Reuse). The results are shown in Table \ref{retrieval}. PKS demonstrates a significant advantage in computational efficiency while maintaining great general performance. The higher performance of PKS compared with PKS w/o Reuse indicates that our reuse strategy not only effectively shortens the prediction time but also improves the performance. More details of variant models are shown in the Appendix \ref{sec:effectofPKS}. Additionally, the ablation study of the influence of molecular fingerprints and graphs is shown in Appendix \ref{sec:ablationPKS}.  

\begin{figure}[!t]\centering
	\includegraphics[width=.98\columnwidth]{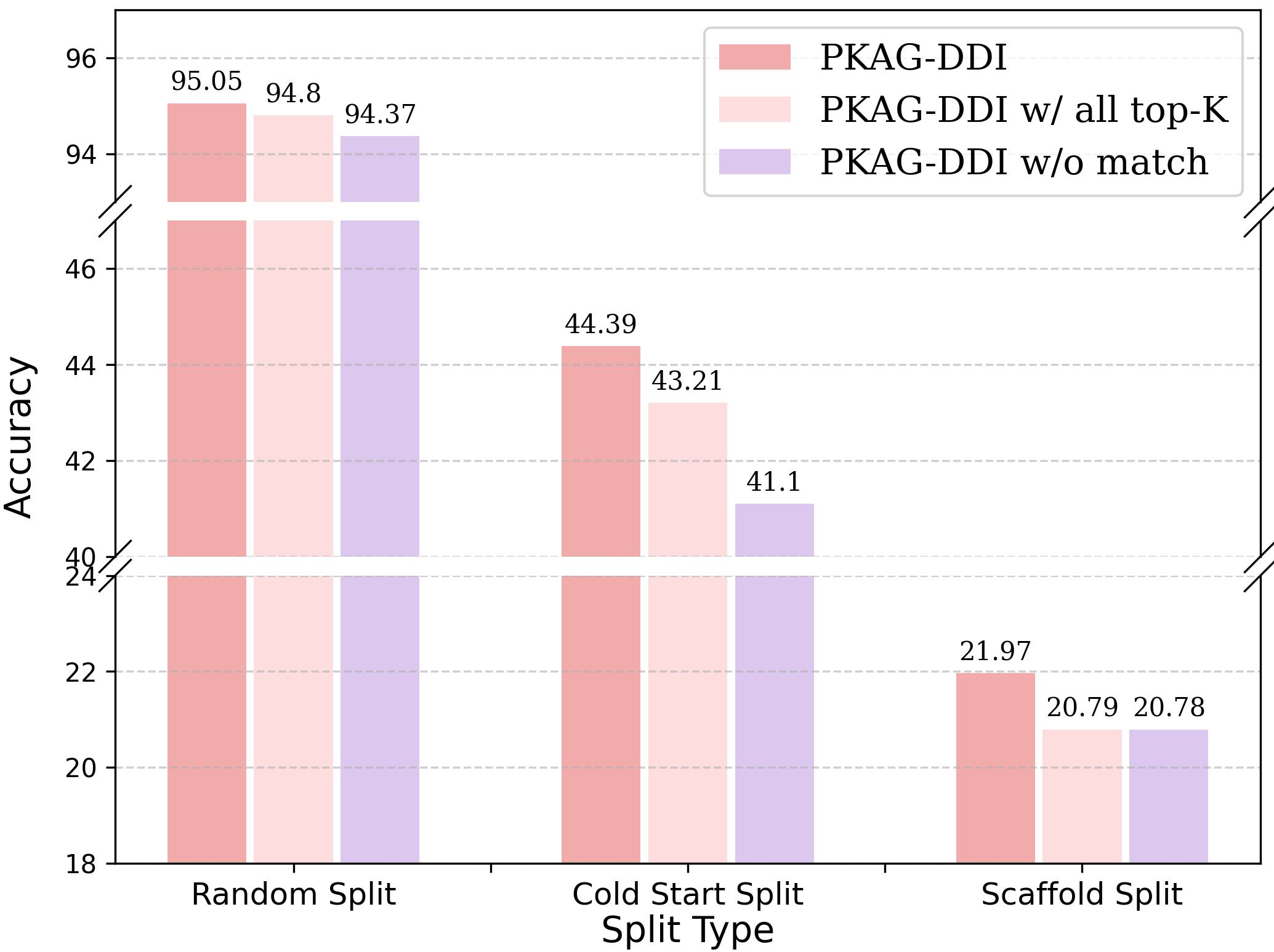}
	\caption{ Results of different integration strategies.}
    \label{ablation}
\end{figure}

\begin{figure*}[!t]\centering
	\includegraphics[width=0.95\textwidth]{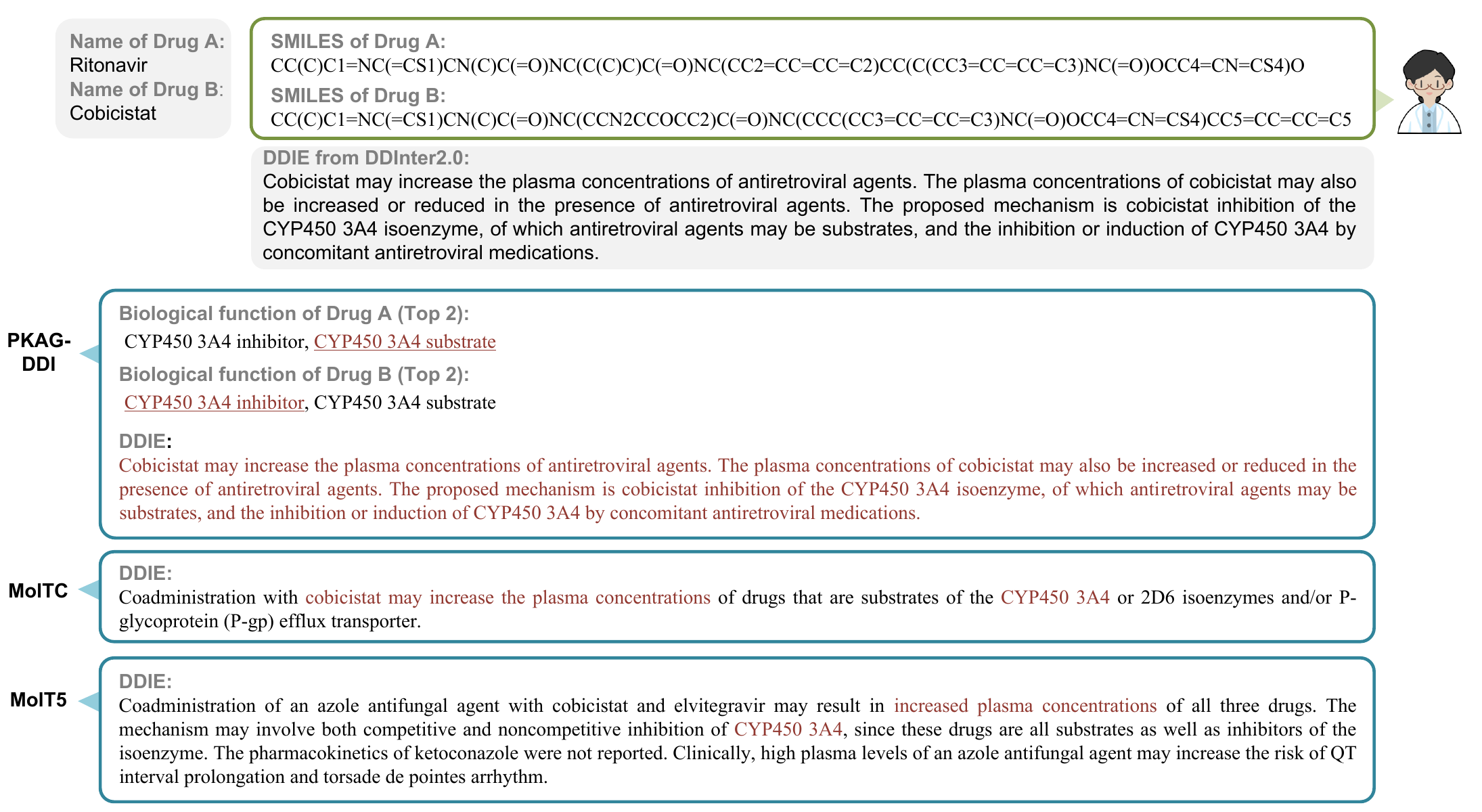}
	\caption{The case study on DDInter2.0. Red text denotes content matching the reference labels. The underline indicates the gold biological function provided by MecDDI.}
    \label{case}
\end{figure*}

\subsection{Effectiveness of Pairwise Knowledge Integration Strategy.}
In this section, we discuss the effectiveness of our proposed pairwise knowledge integration strategy in DDIE generation based on the selected biological function from PKS. We constructed two variants, PKAG-DDI w/ all top-$K$ and PKAG-DDI w/o match. PKAG-DDI w/ all top-$K$ replaces our integration strategy to directly attach all top-$K \times K$ pairwise biological functions to the query input. PKAG-DDI w/o match simply matches pairwise biological functions according to their ranking and without using a joint probability distribution. The DDIE prediction results are shown in Figure \ref{ablation}. It can be observed that PKAG-DDI achieves the best results in all three scenarios, which confirms that our integration strategy can effectively integrate the selected biological functions, thereby maximizing the enhancement of the LM's ability to generate DDIE text. The performance of PKAG-DDI w/ all top-$K$ is inferior to PKAG-DDI, indicating that inputting all potential biological functions may bring noise to LM. PKAG-DDI w/o match achieves the worst performance, which demonstrates that simply matching based on rankings could lead to mismatched biological functions thereby introducing noise, and simply training all instances indiscriminately with the same label may confuse the model.

\subsection{Case Study}

To explicitly show the generation process and the quality of our method, we present a case study in Figure \ref{case}. We choose two CYP3A enzyme inhibitions, Ritonavir and Cobicistat, from \cite{zhong_learning_2024} as the example. When the user inputs the SMILES of the drug pair, our method can explicitly interpret the potential biological functions of each drug and further generate the DDIE text. We find that the gold biological functions are in our prediction, and the DDI text we provide is the same with the reference label, which demonstrates the effectiveness and practical application ability of our method. Alongside completely accurate predictions, we also showcase randomly selected examples containing partial prediction mismatches, which are shown in Appendix \ref{sec:casestudy}.

\section{Conclusion}
In this paper, we emphasized the effectiveness of biological functions in DDIE text generation and introduced a novel pairwise knowledge-augmented generative method for DDIE generation, which can be applied to practical prediction scenarios where knowledge is absent. We also introduce a pairwise knowledge selector to efficiently inject mutually conditional drug information and a pairwise knowledge integration strategy for matching and selectively integrating the knowledge to LM. Experiments demonstrate the superiority of our method over baselines. Our method provides a foundation for transitioning from classification to generation in future DDIE predictions. 

\section*{Limitations}
Although our work has researched the generalization and practical application scenarios, such as inductive sets where test drugs are unseen in the training set, and the case where biological functions are absent, it does not address the zero-shot scenarios where new drugs have novel biological functions that are not included in the existing biological function set. Because the pairwise knowledge selector we provide depends on a fixed amount of knowledge set, and does not support dynamic updating of datasets. Though the variant of PKS (i.e., PKR w/ BERT and PKR w/ BoW) can address the issue, their accuracy and efficiency also need to be improved. Future endeavors will focus on more practical scenarios.

\section*{Acknowledgments}

This work was supported by the National Natural Science Foundation of China (62372204, 62072206, 62102158); Huazhong Agricultural University Scientific \& Technological Self-innovation Foundation; Fundamental Research Funds for the Central Universities (2662024SZ006). The funders have no role in study design, data collection, data analysis, data interpretation, or writing of the manuscript

\bibliography{custom}

\appendix

\section{Preliminary Experiment}
\label{sec:pre}
To evaluate the effectiveness of the biological function in the DDIE prediction task, we experiment two basic DDIE classification models (\textbf{DDI-Finger} and \textbf{DDI-Graph}) and their variants that added the biological function using MecDDI dataset. These two models utilize the popular feature of drug molecules (i.e., molecular 2D graphs and fingerprints \footnote{A widely utilized binary bit vector representing molecular substructures in drug discovery.}) that SMILES can easily convert. Specifically, for DDI-Finger, we employ the RDKit tool to convert SMILES representations into Extended Connectivity Fingerprints (ECFP) with 1024 dimensions for drug molecules. The fingerprints of drug pairs are then concatenated and fed into a 3-layer Multi-Layer Perceptron (MLP) classifier. For DDI-Graph, we utilize the RDKit tool to convert SMILES into 2D molecular graphs. Subsequently, we apply a 3-layer Graph Isomorphism Network (GIN) to obtain the representations of drug pairs, which are concatenated and then input into the 3-layer MLP for classification. Moreover, the variants of the two models are adding different kinds of biological functions. \textbf{DDI-Finger w/ all inherent bio} refers to the DDI-Finger incorporating the representations of drug pairs concatenated with the corresponding one-hot vector of all inherent biological functions of drugs. \textbf{DDI-Finger w/ most relevant bio} means that the model injects the most DDIE-relevant biological function. \textbf{DDI-Graph w/ all inherent bio} and \textbf{Graph w/ most relevant bio} are the same. 

As Figure \ref{toy} shows, the accuracy score of both DDI-Finger and DDI-Graph improves dramatically when injecting the biological function information, suggesting that we can preliminarily conclude that biological functions have the potential to enhance DDI prediction performance and hold value for further research. In addition, we find that the accuracy of DDI-Finger w/ all inherent bio and DDI-Graph w/ all inherent bio is lower than that of DDI-Finger w/ most relevant bio and DDI-Graph w/ most relevant bio, respectively, indicating that the irrelevant biological function in the current DDI may bring noise and hamper the correct prediction. According to this, in this work, we are dedicated to predicting the most DDIE-relevant biological function of drugs in different DDI for better DDIE generation.       

\section{The Discussion of Choosing Gold Biological Function for PKS Training  }
\label{sec:goldbiofunction}

Herein, we discuss the reasons that we chose the drug's gold biological function for multi-class classification prediction in PKS rather than multi-label classification, and the reason that we use BM25 to pick the most similar biological function to DDIE as the gold one. (1) In the MecDDI dataset, only ~19\% of drug pairs have multiple biological functions, and most of these functions are highly similar (e.g., "hyperglycemia" and "hyperglycemic effects"). Adopting multi-label classification solely to accommodate these rare cases could degrade overall prediction accuracy. Since errors in the first stage (biological function prediction) may propagate and amplify in the second stage (DDI generation), we prioritize single-label classification (i.e., choosing a gold label) to maximize prediction accuracy. (2) Given the inherent context sensitivity of autoregressive language models, the more token-similar the input and output text, the stronger the guidance of the input text for the label prediction, thereby improving the accuracy of prediction. Accordingly, for instance, when evaluating a drug's biological functions ("Antihypertensive agent" versus "Hypotensive effects") against its DDIE description ("Pharmacodynamic additive effects (Additive \textbf{hypotensive effects})"), BM25 scoring demonstrates higher similarity for "\textbf{Hypotensive effects}" due to its direct lexical correspondence. Thus, although the “Antihypertensive agent” also shows the semantic relevance with the DDIE, according to the BM25 score, we chose the “Hypotensive effects” as the gold biological function.

\section{Experiment Set }

\subsection{Dataset}
\label{sec:Dataset}
\paragraph{DDIE Text Descriptions.} Considering the increased demand for distinct and high-quality DDI mechanism descriptions, our work focuses on generating more detailed pharmacokinetic and pharmacodynamic event descriptions for DDI. 
Therefore, we use the MecDDI database, which provides drug biological function information, and the DDInter2.0 database, which offers detailed DDI descriptions, for our experimental analysis. Several examples of DDIE text descriptions are shown in Table \ref{descriptions}. Compared with general DDI databases, such as DrugBank, the DDIE from databases MecDDI and DDInter2.0 is more specific. Moreover, the examples illustrate that MecDDI shares more keywords with DDInter2.0 in DDI event descriptions than DrugBank does, which demonstrates that the biological functions from MecDDI may also be applicable to the DDIE prediction in DDInter2.0.

\paragraph{Task Setting} 

In constructing the MecDDI dataset, we find that 99.71\% of drug pairs have only a single DDIE. Consequently, we formulate the DDIE classification in the MecDDI dataset as a multi-class classification task rather than a multi-label classification task.

\begin{table*}[]
\resizebox{\textwidth}{!}{
\begin{tabular}{l|l}
\toprule
 Drug Pair & Mirtazapine \& Ivosidenib \\ \midrule
MecDDI & \begin{tabular}[c]{@{}l@{}}Pharmacodynamic additive effects (Increased risk of prolong QT interval). \end{tabular}\\ \midrule

DDInter2.0 &\begin{tabular}[c]{@{}l@{}}Ivosidenib can cause prolongation of the QT interval. Theoretically, coadministration with other agents \\that can prolong the QT interval may result in additive effects and increased risk of ventricular arrhythmias \\including torsade de pointes and sudden death.\end{tabular} \\ \midrule

DrugBank &\begin{tabular}[c]{@{}l@{}}The metabolism of Mirtazapine can be increased when combined with Ivosidenib.\end{tabular}

 \\ \bottomrule
\end{tabular}}

\resizebox{\textwidth}{!}{
\begin{tabular}{l|l}
\toprule
 Drug Pair & Tenecteplase \& Treprostinil \\ \midrule
MecDDI & \begin{tabular}[c]{@{}l@{}}Pharmacodynamic additive effects (Increased risk of bleeding).\end{tabular}\\ \midrule

DDInter2.0 &\begin{tabular}[c]{@{}l@{}}Drugs that inhibit platelet function may increase the risk of bleeding when administered prior to, during,\\ or after thrombolytic therapy.\end{tabular} \\ \midrule

DrugBank &\begin{tabular}[c]{@{}l@{}}The risk or severity of adverse effects can be increased when Tenecteplase is combined with Treprostinil.\end{tabular}

 \\ \bottomrule
\end{tabular}}

\resizebox{\textwidth}{!}{
\begin{tabular}{l|l}
\toprule
 Drug Pair & Tizanidine \& Opicapone \\ \midrule
MecDDI & \begin{tabular}[c]{@{}l@{}}Pharmacodynamic additive effects (Additive CNS depression effects).  \end{tabular}\\ \midrule

DDInter2.0 &\begin{tabular}[c]{@{}l@{}}The sedative effect of tizanidine may be potentiated by concomitant use of other agents with central nervous \\system (CNS) depressant effects. In addition, tizanidine and many of these agents (e.g., alcohol, anxiolytics, \\sedatives, hypnotics, antidepressants, antipsychotics, opioids, muscle relaxants) also can exhibit hypotensive \\ effects, which may be additive during coadministration and may increase the risk of symptomatic hypotension\\ and orthostasis, particularly during initiation of therapy or dose escalation. Tizanidine itself is a central alpha-2\\ adrenergic agonist. Pharmacologic studies have found tizanidine to possess between 1/10 to 1/50 of the potency\\ of clonidine, a structurally similar agent, in lowering blood pressure.\end{tabular} \\ \midrule

DrugBank &\begin{tabular}[c]{@{}l@{}}The risk or severity of adverse effects can be increased when Tizanidine is combined with Opicapone.\end{tabular}

 \\ \bottomrule
\end{tabular}}

\caption{Examples of the ground-truth DDIE provided by MecDDI, DDInter2.0 and DrugBank, respectively.}
\label{descriptions}
\end{table*}

\subsection{Model Configuration}
\label{sec:config}
PKAG-DDI consists of two stages: biological function selecting and language model learning, which have significant differences in model size. Thus, we used different devices to train them, separately. For the first stage, we developed our model on the machine with a 15 vCPU Intel(R) Xeon(R) Platinum 8362 CPU @ 2.80GHz (CPU) and an NVIDIA GeForce RTX 3090 (GPU). For the second stage, we developed our model on a machine with two A800s with 80GB of video memory. Our model is implemented with PyTorch (2.1.0+cu121), PyTorch-geometric (2.6.1), RDkit (2024.03.5), and pytorch\_lightning (1.9.0). The size of Galactica is 1.3 B. The pre-trained model parameter is the "stage2/last.ckpt" from MolTC.

\subsection{Training Strategy}
In stage two, we employed the AdamW optimizer with a weight decay of 0.05 and a learning rate of 0.001. We implemented an early stopping strategy in the training process to conserve computational resources with "patience" in 5 epochs, and "min\_delta" of training loss is 0.0002. To fair comparison, the MolTC and MolT5 use the same training strategy with our model. Moreover, we fine-tune our model using Low-Rank Adaptation (LoRA) \cite{hu_lora_2021}, which is one of the parameter-efficient fine-tuning (PEFT) technologies.

\subsection{Hyper-parameters}

The primary hyper-parameters, such as the learning rate, weight decay, dropout rate, and the parameter $\lambda$ that controls the information flow from another drug, etc., are searched by using hyper-parameter tuning technology Optuna. The hyper-parameters in PKAG-DDI are represented in the code. Although the biological function pairs grow quadratically with $K$ in theory, in practical applications, most of the biological functions of a drug (around 97\% in the current dataset) are less than three. That is, the larger the $K$, the more noise it will bring and may hinder the DDIE prediction. Thus, we suggest using $K=2$.

\begin{figure}[!t]\centering
	\includegraphics[width=\linewidth]{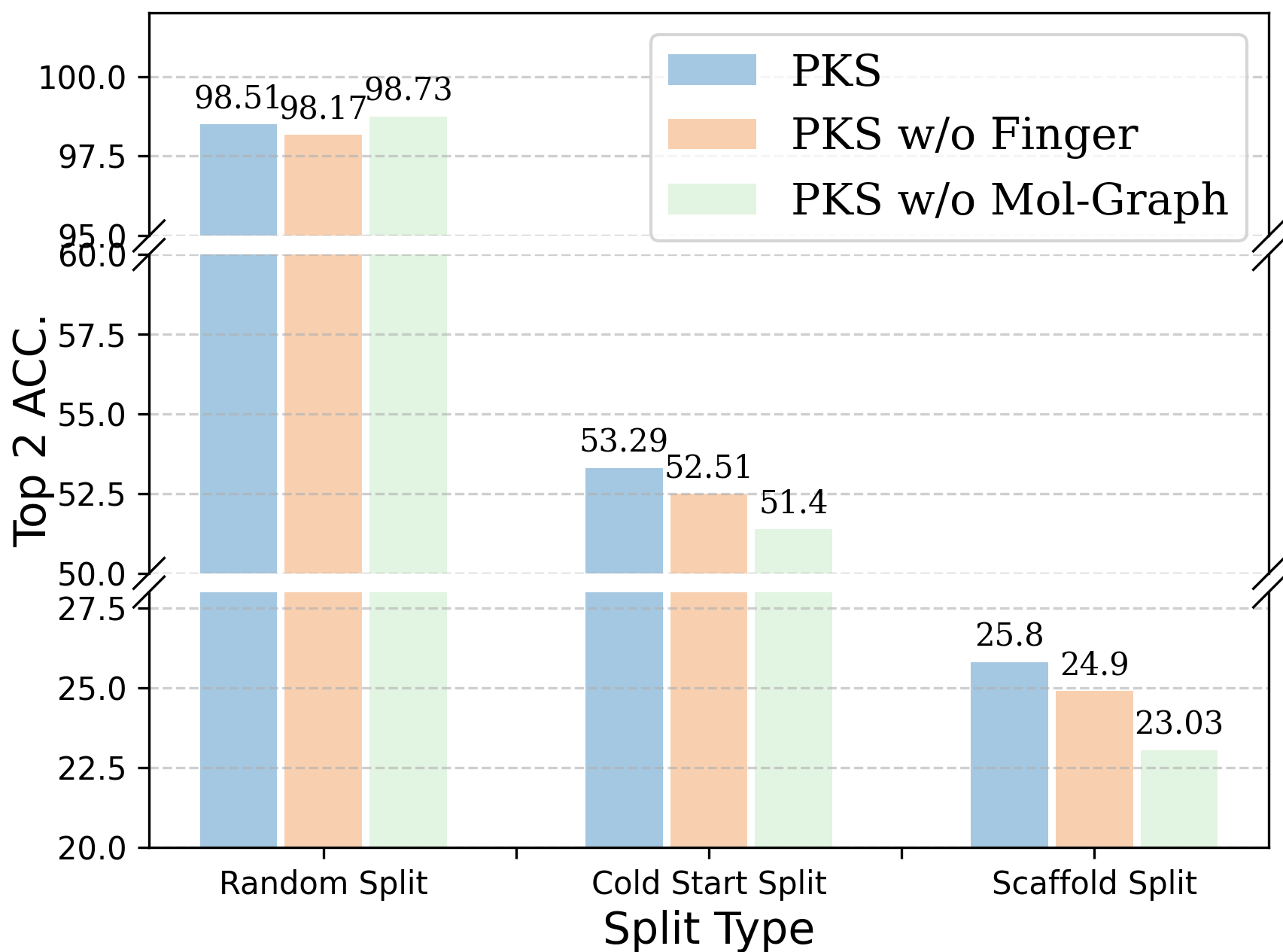}
	\caption{ The ablation study about the information of Fingerprint and Molecular Graph.}
    \label{xiaorong2}
\end{figure}

\begin{figure*}[!t]\centering
	\includegraphics[width=0.98\textwidth]{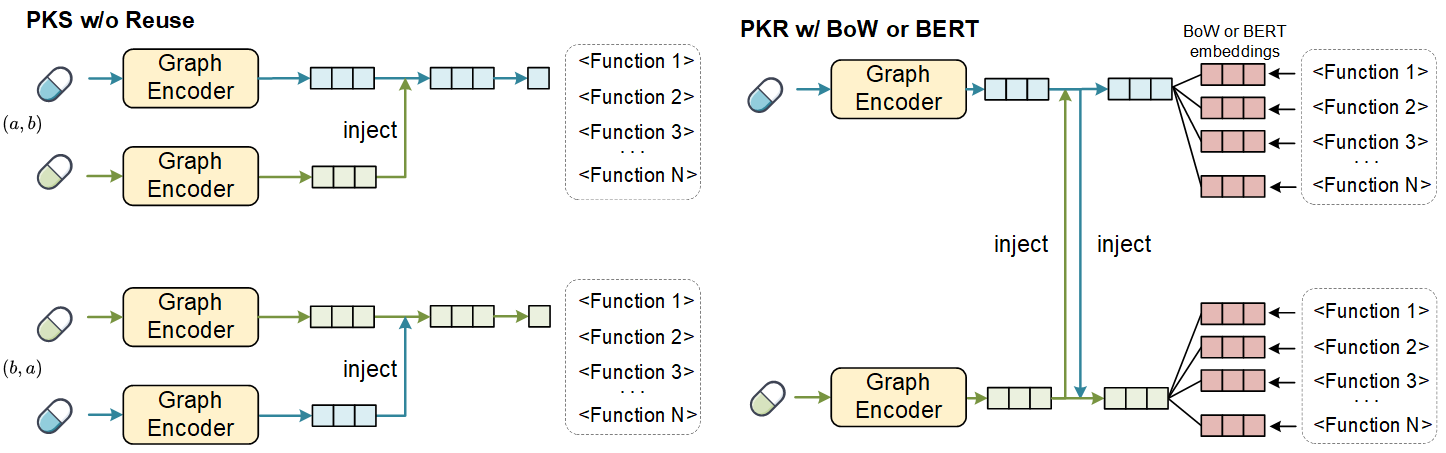}
	\caption{ The variants of PKS.}
    \label{pairwise_retireval}
\end{figure*}

\section{Experiments}

\subsection{Supplement of Efficiency Analysis of PKS}
\label{sec:effectofPKS}
We design two dense retrieval variants of PKS by replacing the classifier with a knowledge encoder and dense inner product model (dubbed pairwise knowledge retriever, PKR). Considering that knowledge (i.e., biological functions) are phrase with an average length of approximately four, we use a common word encoder technology, Bag of Word (BoW), and a semantic encoder technology BERT (with the pre-trained weight \textit{scibert\_scivocab\_uncased}). Moreover, we also design a variant of PKS without the reuse strategy (dubbed PKS w/o Reuse). That is training the instances $(a,b)$ and $(b,a)$ separately. The models' architectures are illustrated in the Figure \ref{pairwise_retireval}. The results shown in Figure \ref{retrieval} demonstrate that PKS outperforms the PKR w/ BoW and PKR w/ BERT in Random Split and Cold Start Split, indicating that directly predicting the label of biological function is enough for most biological function selection scenarios. Additionally, the superior performance of the PKR in the Scaffold Split set indicates that dense retrieval is better suited for scenarios where the test instance distribution differs from the training instance distribution.

\begin{table}[]
\resizebox{\columnwidth}{!}{
\begin{tabular}{ccccc}
\toprule
\multirow{2}{*}{SMILES}  & \multirow{2}{*}{Molecular Graph}  & 
\multicolumn{1}{c}{Random Split} & \multicolumn{1}{c}{Cold Start Split} & \multicolumn{1}{c}{Scaffold Split} \\ 
&& \multicolumn{1}{c}{ACC.}   &  \multicolumn{1}{c}{ACC.}     &
 \multicolumn{1}{c}{ACC.}  

 \\ \midrule
$\checkmark$      &  $\times$ & 94.94 & 43.95 & 21.55 \\
 $\times$  &$\checkmark$  & 94.95& 43.86& 19.98  \\
$\checkmark$ & $\checkmark$ & 	\textbf{95.05} & \textbf{44.39} & \textbf{21.97} \\ \bottomrule  
\end{tabular}}
\caption{The ablation study about evaluating the impact of SMILES and 2D molecular graph in the generator of PKAG-DDI.}
\label{ablation_smiles_mol}
\end{table}

\subsection{Ablation Study of PKS}
\label{sec:ablationPKS}
To evaluate the influence of molecular fingerprints and molecular graphs on PKS. We construct two variants of PKS, PKS w/o Finger and PKS w/o Mol-Graph, by removing molecular fingerprints and graphs, respectively, and compare them with our PKS in three data split scenarios. The results are illustrated in Figure \ref{xiaorong2}. PKS fusing fingerprints and graphs simultaneously achieves the best performance in most scenarios, indicating they help the model comprehensively learn drug molecules and boost the accuracy of biological function selection in more challenging cold and scaffold scenarios. The slightly inferior performance of PKS compared to PKS w/o Mol-Graph in the random split scenario may stem from the fact that when all drugs are seen and samples are abundant, using two types of molecular structural data simultaneously could slightly constrain model optimization.

\subsection{Ablation Study of the Multi-Modality of LM}

In Section 3.3, we use two modalities (that is, the SMILES and the 2D molecular graph) to represent the feature of drugs for LM. Herein, we conduct an ablation study to evaluate the impact of these two modalities on our method. The results shown in Table \ref{ablation_smiles_mol} indicate that the use of both the SMILES information and the molecular graph is beneficial for the generation of DDIE.

\begin{figure*}[!t]\centering
	\includegraphics[width=0.99\textwidth]{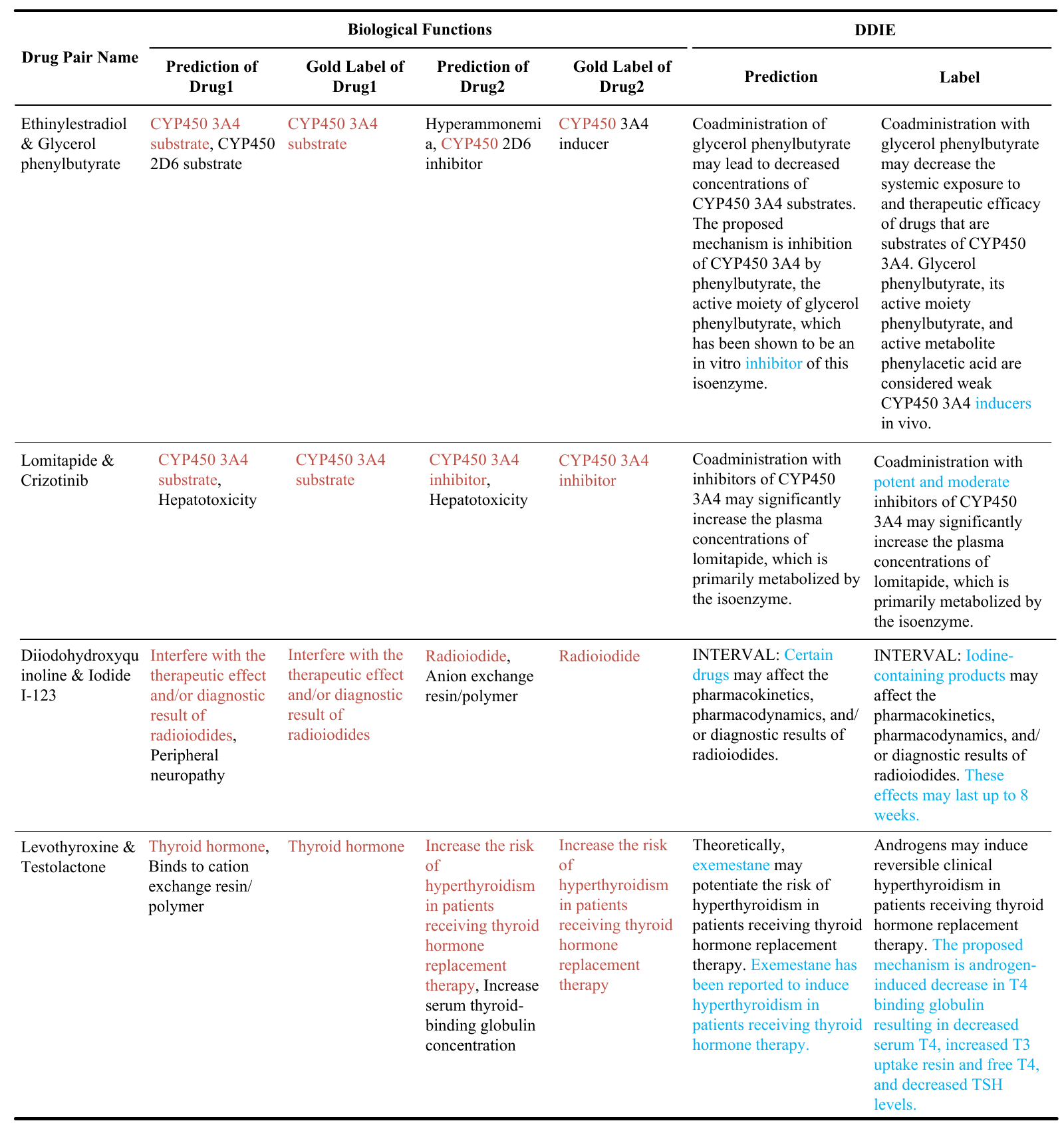}
	\caption{The examples of predictions. The red text indicates matches, while the blue text indicates mismatches.}
    \label{sup_case}
\end{figure*}

% \begin{table*}[]
% \resizebox{\textwidth}{!}{
% \begin{tabular}{lcccccccc}
% \toprule
% \multirow{2}{*}{Split Set} & \multicolumn{4}{c}{First Stage} & \multicolumn{4}{c}{Second Stage} \\
% \cmidrule(lr){2-5} \cmidrule(lr){6-9}
%  & Batch Size & Avg. Time (s) & Max Memory (MB) & Device & Batch Size & Avg. Time (s) & Max Memory (MB) & Device \\
% \midrule
% Random Split & 128 & 65.0 & 22.96 & 1*RTX 3090 & 4 & 7090.4 & 19742.72 & 2*A800 \\
% Cold Split & 128 & 41.9 & 22.56 & 1*RTX 3090 & 4 & 4828.5 & 19640.32 & 2*A800 \\
% Scaffold Split & 128 & 35.2 & 23.63 & 1*RTX 3090 & 4 & 4381.2 & 19230.72 & 2*A800 \\
% \bottomrule
% \end{tabular}}
% \caption{}
% \label{device}
% \end{table*}
\subsection{Supplement of Case Study}
\label{sec:casestudy}
Given PKAG-DDI's superior predictive accuracy (e.g., achieving a 92.39\% BLEU-2), we randomly selected a set of not fully correct test samples from DDInter2.0 to show our method's predictions. As depicted in Figure \ref{sup_case}, we present the Top-2 predictions for biological functions and the predictions for DDIE, along with their corresponding labels. Taking the first DDI sample as an example, PKAG-DDI predicts Glycerol phenylbutyrate as an inhibitor, consequently leading to the DDIE prediction identifying the drug as an inhibitor as well, which is the major difference between this DDIE prediction and its label. This indicates that biological functions have a strong propensity for guiding the generation of DDIEs. Moreover, when the biological functions generated by our method are correct, the resulting DDIE closely matches the labels, such as the second and third examples in Figure \ref{sup_case}.

\end{document}